\documentstyle[aps,prb,epsfig]{revtex}
\begin{document}
\baselineskip=0.5cm
\renewcommand{\thefigure}{\arabic{figure}}
\title{Spin polarization in a two-dimensional electron gas}
\author{B. Davoudi$^{a,b}$ and M. P. Tosi$^a$}
\address{
$^a$NEST-INFM and Classe di Scienze, Scuola Normale Superiore, I-56126 Pisa, 
Italy\\ 
$^b$Institute for Studies in Theoretical Physics and
Mathematics, Tehran, P.O.Box 19395-5531, Iran\\
}
\date{\today}
\maketitle
\begin{abstract}

We evaluate the charge and longitudinal spin response functions of a two-dimensional electron gas with $e^2/r$ interactions in an arbitrary state of spin polarization, using a structurally self-consistent approach to treat exchange and correlations. From the results we assess the nature of the magnetic order in the electronic ground state in zero magnetic field as a function of electron density. We find that states of partial spin polarization are thermodynamically unstable at all values of the coupling strength and that a first-order phase transition occurs with increasing coupling strength from the magnetically disorderd (paramagnetic) phase to the fully spin-polarized (ferromagnetic) phase. This behavior is in qualitative agreement with diffusion Monte Carlo data, although the location of the phase transition is underestimated in our calculations.
\end{abstract}
Pacs: 71.10.Ca; 75.10.Lp
\section{INTRODUCTION}
Electronic systems that exhibit quasi-two-dimensional (2D) behavior have attracted great interest for a number of years. They include electrons in semiconductor heterojunctions and quantum wells and in inversion and accumulation layers at semiconductor-insulator interfaces \cite{Ando}. The basic theoretical model for such systems is the gas of electrons (EG) interacting with the $e^2/r$ law and moving in a plane over a uniform neutralizing background of positive charge.

It is a general feature of interacting electron fluids in both 2D and 3D that spin fluctuations and their correlations become more important as the coupling strength increases with decreasing electron density. This fact has been most strikingly emphasized by quantum simulation work. In the 3D fluid these studies \cite{Ceperly,Ortiz} have revealed a continuous transition from the paramagnetic to the ferromagnetic state, before a first-order transition to a ferromagnetic Wigner crystal occurs. In the 2D EG, instead, similar studies \cite{Tanatar,Rapisarda} have found no evidence for partially spin-polarized ground states, but have indicated that a first-order transition occurs from the paramagnetic to the ferromagnetic fluid before crystallization.

A theoretical study of the magnetic phase behavior of the 2D electron fluid is the purpose of the present work. Previous calculations of charge and spin correlations in the 2D EG \cite{Jonson} have referred to the paramagnetic state only. We consider instead, at a chosen set of values for the Coulomb coupling strength, a 2D EG in an arbitrarily chosen state of spin polarization and evaluate the whole matrix of dynamic susceptibilities for its charge and longitudinal spin response. From this viewpoint we can assess both the absolute dynamic stability and the relative thermodynamic stability of the various states of spin polarization. The basic approximation underlying our calculations is an immediate extension of the so-called STLS approach \cite{Singwi} to treat exchange and correlations through self-consistency between linear susceptibilities and pair distribution functions. This is a necessary starting point to the problem at hand and allows, as we shall see, qualitative understanding to be gained on the magnetic phase behavior of the 2D EG.

The contents of the paper are briefly as follows. Section \ref{Theory} presents the essential definitions of linear response functions and their expressions through local field factors allowing in an unrestricted way for exchange and short-range correlations between the three types of electron pairs in a spin-polarized state \cite{Marinescu}. The approximate expressions of the local field factors embodying self-consistency with the liquid-structure pair functions are also given there. Section \ref{Numerical} reports the main results of the self-consistent calculation of structure and response and then focuses on the numerical results for the ground state energy as a function of electron density and spin polarization. We show that, starting from the usual paramagnetic ground state in the weak-coupling regime where we find very good agreement with the accurate results reported by Yarlagadda and Giuliani \cite{Yarlagadda} for the spin susceptibility, the states of partial spin polarization are squeezed out before a crossing occurs between the energy curves of the paramagnetic and ferromagnetic states. The transition between these two states is found to occur, however, at a lower value of the coupling strength than indicated by the quantum simulations. Finally Section \ref{SAFD} summarizes our main conclusions and gives suggestions for further work.

\section{THEORETICAL APPROACH}\label{Theory}
We consider a 2D EG modelling a system of electronic carriers with effective mass $m^*$ in a semiconductor with background dielectric constant $\epsilon_0$ and interacting {\it via} the Coulomb potential $v_q=2\pi e^2/\epsilon_0 q$. The thermodynamic state parameters of the EG at zero temperature are taken as the areal densities $n_\uparrow$ and $n_\downarrow$ of spin-up and spin-down electrons or equivalently the total density $n=n_\uparrow+n_\downarrow$ and the spin polarization $\zeta=(n_\uparrow-n_\downarrow)/n$. The coupling strength parameter is $r_s=(\pi n {a_B^*}^2)^{-1/2}$, where $a_B^*=\hbar^2 \epsilon_0/m^* e^2$ is the effective Bohr radius.

The charge and longitudinal spin response of the EG to an external electromagnetic field is described in the linear regime by a set of susceptibilities $\chi_{\sigma\sigma'}(q,\omega)$, the subscripts $\sigma$ and $\sigma'$ being spin indices ($\uparrow\,{\rm or}\,\downarrow$) \cite{Marinescu}. These are defined through the double Fourier transform of the linear response functions of the EG to a position and time dependent electromagnetic field, which are given by

\begin {equation}
K_{\sigma\sigma'}(|{\bf r}-{\bf r'}|,t-t')=-i\hbar\, \theta(t-t')\left\langle [\rho_{\sigma}({\bf r},t),\rho_{\sigma'}({\bf r'},t')]\right\rangle.
\end{equation}
Here, $\theta(t)$ is the Heaviside step function and $\rho_{\sigma}({\bf r},t)$ is the spin density operator in the Heisenberg representation. Linear combination of the $\chi_{\sigma\sigma'}$ susceptibilities give the charge susceptibility $\chi_{cc}$ and the spin susceptibility $\chi_{ss}$ as well as the off-diagonal susceptibilities $\chi_{cs}$ and $\chi_{sc}$. Specifically, leaving implicit the dependence on $q$ and $\omega$, we have $\chi_{cc}=e^2\sum_{\sigma\sigma'}\chi_{\sigma\sigma'}$, $\chi_{ss}=-\gamma^2\sum_{\sigma\sigma'}{\rm sgn} (\sigma\sigma')\chi_{\sigma\sigma'}$, $\chi_{cs}=e\gamma\sum_{\sigma\sigma'}{\rm sgn} (\sigma')\chi_{\sigma\sigma'}$ and $\chi_{sc}=e\gamma\sum_{\sigma\sigma'}{\rm sgn} (\sigma)\chi_{\sigma\sigma'}$, with $\gamma=g\mu_B/2$ where $g$ is the Land\'e factor and $\mu_B$ the Bohr magneton. 

The partial structure factors $S_{\sigma\sigma'}(q)$ are related to the linear susceptibilities by the fluctuation-dissipation theorem, which can be written in the form
\begin {equation}\label{Sij}
S_{\sigma\sigma'}(q)=-\frac{\hbar}{\sqrt{n_\sigma n_{\sigma'}}}\int_0^\infty \frac{du}{\pi}
\chi_{\sigma\sigma'}(q,iu)\, .
\end {equation}

The integral in Eq. (\ref{Sij}) is taken along the imaginary frequency axis, the contribution from the plasmon pole being easily handled numerically by this method (see de Freitas {\it et al.}\cite{Jonson}). The pair distribution functions $g_{\sigma\sigma'}(r)$ are obtained from Eq. (\ref{Sij}) by Fourier transform,
\begin{equation}\label{gii}
g_{\sigma\sigma'}(r)=1+\frac{1}{\sqrt{n_\sigma n_{\sigma'}}}\int \frac{d^2 q}{(2\pi)^2}[S_{\sigma \sigma'}(q)-\delta_{\sigma\sigma'}]
\exp(i{\bf q}\cdot{\bf r})
\end{equation}
and the ground state energy $E(r_s,\zeta)$ as a function of electron density and spin polarization can in turn be calculated from these structural functions. The results is
\begin{equation}\label{E}
E(r_s,\zeta)=\frac{1+\zeta^2}{r_s^2}+\frac{\sqrt{2(1+\zeta)}}{r_s^2}\int_0^{r_s} d r_s'\gamma(r_s',\zeta),
\end{equation}
in units of the effective Rydberg ${\rm Ryd}^*=e^2/2\epsilon_0 a^*_B$. The first term in Eq. (\ref{E}) is the free-electron kinetic energy and the second term is the exchange-correlation energy $E_{xc}$, with $\gamma$ defined by
\begin{equation}
\gamma(r_s,\zeta)=\int_0^\infty dq [S_{cc}(q)-1]
\end{equation}
and $S_{cc}(q)=[n_\uparrow S_{\uparrow\uparrow}(q)+n_\downarrow S_{\downarrow\downarrow}(q)+2 (n_\uparrow n_\downarrow)^{1/2}S_{\uparrow\downarrow}(q)]/n$ being the charge-charge partial structure factor.
One can extract the correlation energy $E_c$ from $E_{xc}$ by subtracting the Hartree-Fock exchange part. We have
\begin{equation}
E_c=E_{xc}+\frac{4\sqrt{2}}{3\pi r_s}[(1+\zeta)^{3/2}+(1-\zeta)^{3/2}].
\end{equation}

The local field factors $G_{\sigma\sigma'}(q,\omega)$ are introduced by writing an effective one-electron Hamiltonian in which an electron with spin $\sigma$ experiences effective fields embodying exchange and correlation with the surrounding EG \cite{Marinescu}. In the following we adopt an STLS-like approximation, in which the bare electron-electron interaction $v_q$ is replaced by an effective spin-dependent static interaction $\psi_{\sigma\sigma'}(q)=v_q[1-G_{\sigma\sigma'}(q)]$. The corresponding expression for the linear susceptibilities is
\begin{eqnarray}\label{Xij}
\chi_{\sigma\sigma'}(q,\omega)=\frac{\chi_{0\sigma}(q,\omega)\left [ \delta_{\sigma\sigma'}
+\eta_{\sigma\sigma'}\psi_{\bar{\sigma}\bar{\sigma}'}(q)\chi_{0\bar{\sigma}'}(q,\omega)\right ]} 
{\Delta(q,\omega)}\, ,
\end{eqnarray}
where $\bar{\sigma}$ denotes the spin orientation opposite to $\sigma$, $\eta_{\sigma\sigma'}=(-1)^{\delta_{\sigma\sigma'}}$, $\chi_{0\sigma}(q,\omega)$ are the (diagonal) susceptibilities of the ideal 2D Fermi gas and $\Delta(q,\omega)$ is given by
\begin {eqnarray}\label{Del}
\Delta(q,\omega)=\left[1-\psi_{\uparrow\uparrow}(q)\chi_{0\uparrow}(q,\omega)\right ]
\left[1-\psi_{\downarrow\downarrow}(q)\chi_{0\downarrow}(q,\omega)\right]
-\psi_{\uparrow\downarrow}(q)\psi_{\downarrow\uparrow}(q)\chi_{0\uparrow}(q,\omega)\chi_{0\downarrow}(q,\omega)
\,.
\end {eqnarray}
The expressions for the real and imaginary part of the ideal-gas susceptibility are well known (see e.g. Ref. [\onlinecite{Ando}]). From these the expression for $\chi_{0\sigma}(q,iu)$ on the imaginary frequency axis is obtained for use in Eq. (\ref{Sij}) as
\begin{eqnarray}\label{X0i}
\chi_{0\sigma}(q,iu)=\frac{{m^*}^2}{2\pi \hbar^2q^2} \left[\sqrt{2}\sqrt{a_\sigma+\sqrt{a_\sigma^2+(\frac{q^2u}{\hbar m^*})^2}}-\frac{q^2}{m^*}\right]
\end{eqnarray}
where we have defined 
\begin{equation}
a_\sigma=\frac{q^4}{4{m^*}^2}-\frac{q^2k_{F\sigma}^2}{{m^*}^2}-\frac{u^2}{\hbar^2}
\end{equation}
with  $k_{F\sigma}=(4\pi n_\sigma)^{1/2}$. Finally, the local field factors are determined from the partial structure factors through the closure relation \cite{Singwi}
\begin {equation}\label{Gij}
G_{\sigma\sigma'}(q)=-\frac{1}{\sqrt{n_\sigma n_{\sigma'}}}\int \frac{d^2k}{(2 \pi)^2}
\frac{q}{k}[S_{\sigma\sigma'}(|{\bf k}-{\bf q}|)-\delta_{\sigma\sigma'}].
\end {equation}
A self-consistency cycle can thus be set up by using Eq. (\ref{Xij}) in Eq. (\ref{Sij}), leading at convergence to numerical results for the partial structure factors and for the local field factors.

With the ground state energy as a function of $r_s$ and $\zeta$ at our disposal we can calculate the compressibility and the spin susceptibility of the EG. The compressibility $\kappa$ is given by the partial derivatives of the ground state energy with respect to the density:
\begin{equation}\label{K1}
\frac{1}{n\kappa}=\left. n\frac{\partial P}{\partial n}\right|_\zeta=\frac{1}{4}\left(\left. r_s^2\frac{\partial^2E}{\partial r_s^2}\right|_\zeta-\left. r_s\frac{\partial E}{\partial r_s}\right|_\zeta\right)
\end{equation}
where we have used the definition of the pressure $P=-(nr_s/2)\partial E/\partial r_s|_\zeta$. 
The spin susceptibility $\chi_s\equiv\lim_{q\rightarrow 0}\chi_{ss}(q,0)$ can similary be obtained from the second derivative of the ground state energy with respect to magnetization:
\begin{equation}\label{Xs1}
\frac{\chi_P}{\chi_s}= \frac{m^*}{\pi n\hbar^2}\left.\frac{\partial^2E}{\partial\zeta^2}\right|_{r_s}.
\end{equation}
The quantity $\chi_P=m^*g^2\mu_B^2/4\pi\hbar^2$ is the Pauli spin susceptibility of the ideal 2D Fermi gas.
An alternative route to the compressibility and the spin susceptibility is through the long-wavelenght limit of the local field factors \cite{Marinescu}, as will be discussed for the paramagnetic state in Section III. B below.

The function $\Delta(q,\omega)$ also carries valuable information about the collective excitations and the dynamical stability of the system. The plasmon dispersion relation $\omega_p(q)$ is found by solving the equation $\Delta(q,\omega)=0$. The existence of a solution for the equation $\Delta(q,\omega)=0$ when $\omega$ is purely imaginary is a signature of a dynamical instability in the system. We shall meet this instability in Section III. C.

As a final remark, the matrix of charge-spin susceptibilities becomes diagonal in the paramagnetic state, yielding
\begin{eqnarray}\label{Xcs}
\chi_{(cc,ss)}(q,\omega)\propto\frac{\chi_0(q,\omega)}{1-\frac{1}{2}[\psi_{\uparrow\uparrow}(q)\pm\psi_{\uparrow\downarrow}(q)]\chi_0(q,\omega)}.
\end{eqnarray}
where $\chi_0(q,\omega)=2\chi_{0\uparrow}(q,\omega)=2\chi_{0\downarrow}(q,\omega)$.  Equations (\ref{K1}) and (\ref{Xs1}) can be rewritten as
\begin{equation}
\frac{\kappa_0}{\kappa}=1-\frac{\sqrt{2}r_s}{\pi}+\frac{r_s^4}{8}\left[\frac{\partial^2E_c}{\partial r_s^2}-\frac{1}{r_s}\frac{\partial E_c}{\partial r_s}\right]_{\zeta=0}
\end{equation}
and
\begin{equation}
\frac{\chi_P}{\chi_s}=1-\frac{\sqrt{2}}{\pi}r_s+\frac{r_s^2}{2}\left[\frac{\partial^2E_c}{\partial \zeta^2}\right]_{\zeta=0},
\end{equation}
where $\kappa_0=\pi r_s^4/2$ is the ideal-gas compressibility in units of ${a_B^*}^2/Ryd^*$.
\section{NUMERICAL RESULTS}\label{Numerical}
In this section we report our numerical results for the structure and local field functions and for the ground-state energy over relevant ranges of the coupling-strength parameter $r_s$ and of the magnetization fraction $\zeta$. We also report the plasmon dispersion relation and discuss the stability of the 2D EG as a function of $\zeta$.

\subsection{CORRELATION FUNCTIONS AND LOCAL FIELDS}
Figures \ref{Fig1} and \ref{Fig2} show the behavior of the partial structure factors as functions of $q/k_{F\uparrow}$, first with varying $\zeta$ at $r_s=2$ and then with varying $r_s$ at $\zeta=0.5$. Starting from the left in each figure, the three sets of curves give $S_{\downarrow\downarrow}(q)$, $S_{\uparrow\uparrow}(q)$ and $S_{\uparrow\downarrow}(q)=S_{\downarrow\uparrow}(q)$. 

It is seen from Figure \ref{Fig1} that with increasing $\zeta$ away from the paramagnetic state ({\it i.e.} with the induction of spin polarization in the up direction) the function $S_{\uparrow\uparrow}(q/k_{F\uparrow})$ remains essentially the same while $S_{\downarrow\downarrow}(q/k_{F\uparrow})$ shrinks towards the vertical axis. These behaviors descend from the fact that these structural functions mainly reflect the Pauli exchange hole between parallel-spin electrons. Accordingly , these structure factors are almost independent of the Coulomb coupling strength, as is seen from Figure \ref{Fig2}. The cross structure factor $S_{\uparrow\downarrow}(q/k_{F\uparrow})$, on the other hand, shows a trough due to correlations from Coulomb repulsions, which deepens with increasing coupling strength in Figure \ref{Fig2}. The shift of the trough of $S_{\uparrow\downarrow}(q/k_{F\uparrow})$ in Figure \ref{Fig1}, as well as the shrinking of 
$S_{\downarrow\downarrow}(q/k_{F\uparrow})$, are mainly due to the increase of the scaling wave numbers $k_{F\uparrow}$ with increasing $\zeta$ at constant total electron density.

The behavior of the Pauli exchange and Coulomb holes is directly seen in the pair distribution functions $g_{\sigma\sigma'}$ and is illustrated with varying $\zeta$ at $r_s=2$ in Figure \ref{Fig3} and with varying $r_s$ at $\zeta=0.5$ in Figure \ref{Fig4}. The holes in $g_{\uparrow\uparrow}(r)$ and $g_{\downarrow\downarrow}(r)$ show little dependence on both coupling strength and spin polarization (except for scaling of the unit of length). They do not exactly vanish as $rk_{F\uparrow}\rightarrow 0$, which evidently is a quantitative defect of the present approximate approach. The Coulomb hole in $g_{\uparrow\downarrow}(r)=g_{\downarrow\uparrow}(r)$, on the other hand, shows appreciable dependence on coupling strength (as expected) and on spin polarization. Except for the region of small $rk_{F\uparrow}$, an increase in spin polarization broadens the Coulomb hole and thus brings about an effective decrease in the Coulomb repulsive energy. This ultimately leads to stabilization of the ferromagnetic state. The very weak dependence of $g_{\uparrow\downarrow}(0)$ on the state of spin polarization has already been discussed elsewhere \cite{Polini}.

Finally, Figure \ref{Fig5} shows that the local field factors $G_{\sigma\sigma'}(q)$ have a rather weak dependence on spin polarization at fixed $r_s=2$. The asymptotic values of these functions for $q\rightarrow\infty$ are determined by the relations $G_{\uparrow\uparrow}(\infty)=G_{\downarrow\downarrow}(\infty)=1$ and $G_{\uparrow\downarrow}(\infty)=1-g_{\uparrow\downarrow}(0)$ \cite{Marinescu}: these relations are rather well satisfied in our approximate approach. The slopes of $G_{\sigma\sigma'}(q)$ in the long wavelength limit are instead related to the compressibility and spin susceptibility and will be discussed below.

\subsection{GROUND STATE ENERGY}\label{GSE}

We evaluate the ground state energy $E$ as a function of $r_s$ and $\zeta$ from Eq. (\ref{E}). Figure \ref{Fig6} reports $E$ as a function of $r_s$ for various values of $\zeta$ in the range 0 to 1. The paramagnetic state has the lowest ideal-gas kinetic energy and remains thermodynamically stable over the whole range of coupling strength illustrated in Figure \ref{Fig6}, up to $r_s=4$. The energy difference from the ferromagnetic state is steadily decreasing and the energy curves corresponding to intermediate values of $\zeta$ are being squeezed together above their various minima. In fact, as we shall show in full detail in the next section, the states with partial spin polarization become dynamically unstable in the upper part of this range of $r_s$. Only the paramagnetic and fully spin polarized states remain dynamically stable above $r_s\cong 4$ and their energy curves cross at $r_s=5.5$, as is seen from Figure \ref{Fig7}.

The present approximate approach thus predicts a first-order quantum phase transition occurring from the paramagnetic to the ferromagnetic state at $r_s=5.5$ in the 2D EG at zero temperature. Above their crossing, the energy curves of the two states remain very close to each other over the whole range of $r_s$ illustrated in Figure \ref{Fig7}, up to $r_s=10$. There is, therefore, a large uncertainty attached to our prediction of the location of the phase transition.

Figure \ref{Fig8} reports a number of different results for the spin susceptibility $\chi_s$ of the 2D EG in the paramagnetic state, in units of the ideal Pauli susceptibility $\chi_P$, as a function of $r_s$ up to $r_s=4$. The values of $\chi_s/\chi_P$ that we obtain by differentiation of the ground state energy from Eq. (\ref{Xs1}) (denoted as STLS2 in Figure \ref{Fig8}) are in reasonable agreement with the values extracted from the available quantum Monte Carlo data \cite{Rapisarda}, which are shown as crosses in Figure \ref{Fig8}. We also obtain close agreement with the theoretical results of Yarlagadda and Giuliani \cite{Yarlagadda} at low values of $r_s$, although these values of $\chi_s/\chi_P$ increase much too rapidly with coupling strength beyond the range of $r_s$ studied in the original work of these authors. A general comment applying to both their results and ours is that the role of exchange is somewhat overemphasized relative to the Monte Carlo data. The location of the phase transition is accordingly placed at a lower $r_s$, as we have already seen in Figure \ref{Fig7}.

There is, however, a large violation of the susceptibility sum rule in our approach: the value of $\chi_s/\chi_P$ that we obtain from the slopes of the local field factors at long wavelengths according to 
\begin{equation}\label{Xs2}
\frac{\chi_p}{\chi_s}=1-\lim_{q\rightarrow 0}[G_{\uparrow\uparrow}(q)+G_{\downarrow\downarrow}(q)-G_{\uparrow\downarrow}(q)-G_{\downarrow\uparrow}(q)]/qa_B^*
\end{equation}
(denoted as STLS1 in Figure \ref{Fig8}) are vastly different from those obtained from the ground state energy according to Eq. (\ref{Xs1}) and indeed their dependence on $r_s$ is much too close to that following from the simple Hartree-Fock approximation.

Finally, the results that we obtain for the compressibility of the paramagnetic state by density differentiation of the ground state energy according to Eq. (\ref{K1}) are in excellent quantitative agreement with the Monte Carlo data \cite{Tanatar,Rapisarda}. There is again a large violation of the compressibility sum rule, this being a well known defect of the STLS approach \cite{Singwi}.

\subsection {PLASMON DISPERSION AND DYNAMIC INSTABILITY OF PARTIALLY POLARIZED STATES}\label{PADI}
The plasmon excitation energy $\omega_p(q)$ is evaluated from the root of the equation $\Delta(q,\omega_p(q))=0$, $\Delta(q,\omega)$ being as given in Eq. (\ref{Del}). The results are shown in Figure \ref{Fig9} at $r_s=2$ for various values of the spin polarization, superposed on the particle-hole continuum for the majority spin population. It can be shown analytically from Eq. (\ref{Del}) that the plasmon dispersion relation at long wavelength is given by $\omega_p(q)\rightarrow \sqrt{2\pi (n_\uparrow+n_\downarrow)e^2 q/\epsilon_0 m^*}$.

We turn next to examine the behavior of the function $\Delta(q,\omega)$ on the imaginary axis $\omega=i u$. We notice first that the energy change $\Delta E$ associated with a small modulation $\delta n_\sigma({\bf q})$ of the spin populations away from a homogeneous state with densities $n_\uparrow$ and $n_\downarrow$ is given by
\begin{eqnarray}\label{DE}
\Delta E=-\sum_{\sigma,\sigma'}\int \frac{d^2q}{(2\pi)^2}\chi^{-1}_{\sigma\sigma'}(q)\delta n_{\sigma}({\bf q}) \delta n_{\sigma'}({\bf q}).
\end{eqnarray}
This quantity must be positive if the undeformed spin state is stable and, using Eq. ({\ref{Xij}) and the fact that $\chi_{0\sigma}(q,0)$ is negative for all values of q, we find that the stability condition can be written as
\begin{equation}\label{ST}
\Delta(q,0)>0.
\end{equation}
In fact, it can be shown analytically from Eq. (\ref{Del}) that in the paramagnetic state the stability condition in Eq. (\ref{ST}) reduces at long wavelengths to the condition that the spin susceptibility defined in Eq. (\ref{Xs2}) be positive.

Figure \ref{Fig10} reports the function $\Delta(q,iu)$ plotted against $u$ at $\zeta=0.5$ and various values of $r_s$ for a small value of $q$ ($q=0.1 k_{F\uparrow}$). It is evident that this state of partial spin polarization is becoming dynamically unstable at $r_s\approx3$, as signalled by a change in sign of the function $\Delta(q,iu)$ at low wave number and frequency.

In view of the violation of the susceptibility sum rule in the present theory as discussed in Section III. B we should, however, return to Figures \ref{Fig6} and \ref{Fig7} and assess the thermodynamic stability of the spin-polarization states for the 2D EG by the well tested and reliable results reported these. That is, states of partial spin polarization are never stable and the system goes from paramagnetic to ferromagnetic by a first-order phase transition.

As a final remark, we find no dynamic instability in the charge-charge response function of the 2D EG in either the paramagnetic or the ferromagnetic state over a wide range of $r_s$ (up to $r_s=30$).

\section{SUMMARY AND FUTURE DIRECTIONS}\label{SAFD}
In summary, we have evaluated the linear charge and longitudinal spin susceptibilities of a 2D electron gas in various states of spin polarization over a range  of electron densities, using a structurally selfconsistent approach to treat exchange and correlations. The main results of the calculations are the structural pair functions and hence the energy of the system as a function of electron density and spin polarization. These results have in particular been tested for the paramagnetic state against previous theoretical results and quantum Monte Carlo simulations of the spin susceptibility and of the compressibility.

From the calculated energy function we have found that the 2D electron gas undergoes a first-order phase transition from the paramagnetic to the ferromagnetic state with increasing coupling strength, without passing through intermediate states of partial magnetization. This result agrees with the scenario coming from quantum Monte Carlo simulations, although our predicted location of the phase transition appears to be underestimated.

We have stressed that the main defect of our approach is the lack of thermodynamic consistency in the compressibility and spin susceptibility, {\it i.e.} the disagreement between the values obtained by differentiation from the energy function and those derived from the appropriate limit of the microscopic susceptibilities. The latter show dynamic instabilities emerging in the states of partial magnetization as the coupling strength increases. These inconsistencies point the way to possible improvements in the theoretical approach through a refinement of the closure relation in Eq. (\ref{Gij}). As is suggested by early work on the 3D paramagnetic electron gas \cite{Singwi,Vashishta}, thermodynamic consistency may be imposed by allowing for the density and spin-polarization dependence of the partial structure factors used in the evaluation of the local fields in the microscopic susceptibilities.
\acknowledgements
{
This work was partially supported by MURST through the PRIN1999 program. We thank Dr. M. Polini for useful discussions.
}

\begin{figure}
\centerline{\mbox{\psfig{figure=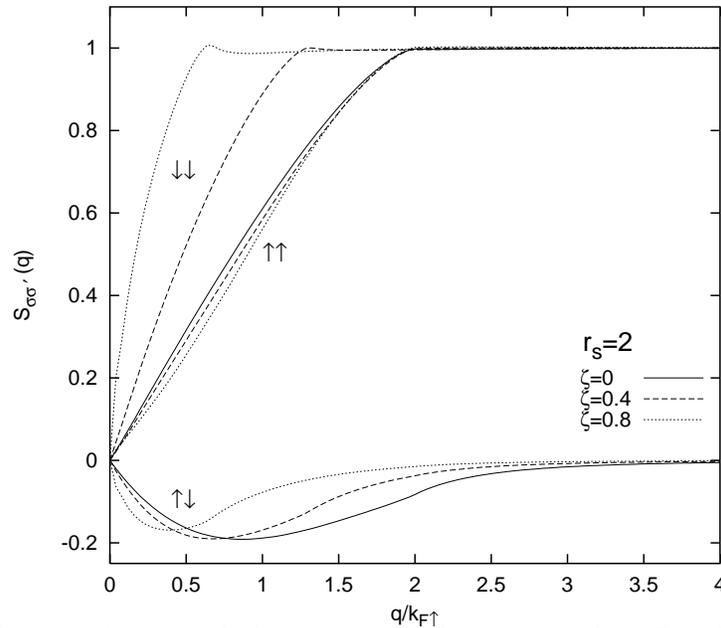, angle =0, width =10 cm}}} 
\caption{Static structure factors as functions of $q/k_{F\uparrow}$ at $r_s=2$ and $\zeta=0\, , 0.4\, {\rm and}\, 0.8$. From left to right: $S_{\downarrow\downarrow}(q)$, $S_{\uparrow\uparrow}(q)$ and $S_{\uparrow\downarrow}(q)$.}
\label{Fig1}
\end{figure}

\begin{figure}
\centerline{\mbox{\psfig{figure=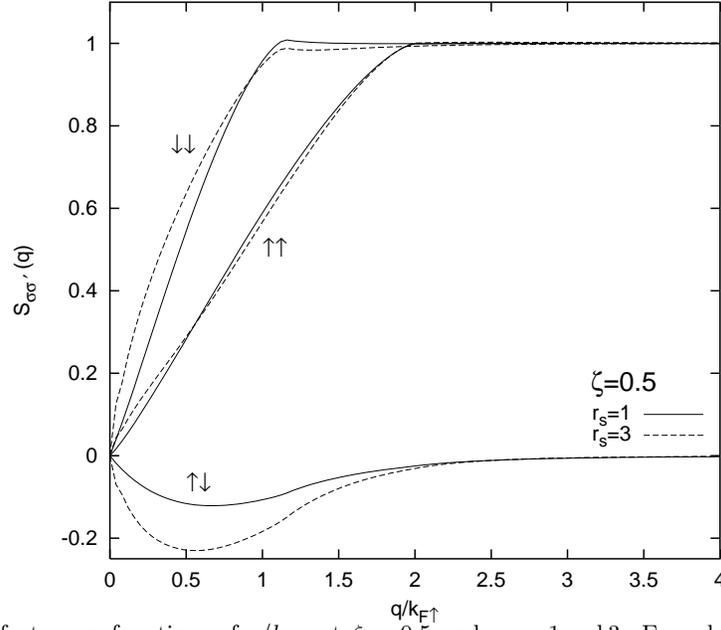, angle =0, width =10 cm}}} 
\caption{Static structure factors as functions of $q/k_{F\uparrow}$ at $\zeta=0.5$ and $r_s=1\, {\rm and}\,3$. From left to right: $S_{\downarrow\downarrow}(q)$, $S_{\uparrow\uparrow}(q)$ and $S_{\uparrow\downarrow}(q)$.}
\label{Fig2}
\end{figure}

\begin{figure}
\centerline{\mbox{\psfig{figure=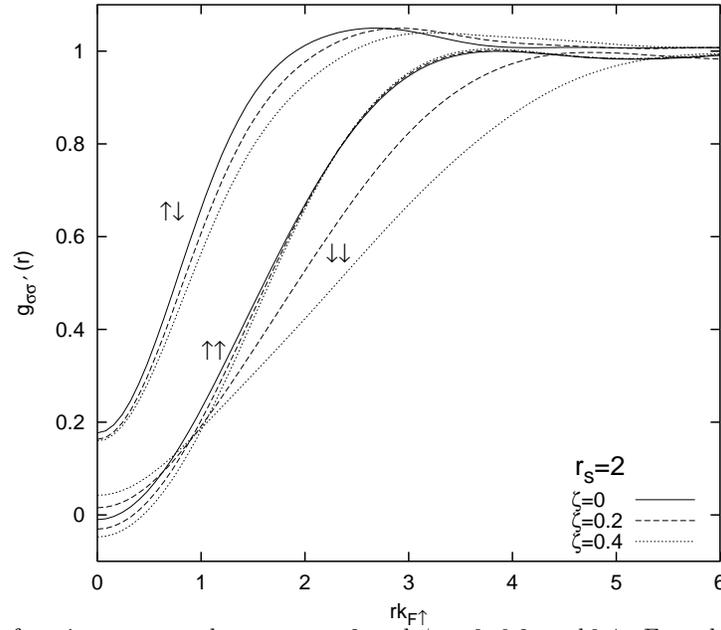, angle =0, width =10 cm}}} 
\caption{Pair distribution functions {\it versus} $rk_{F\uparrow}$ at $r_s=2$ and $\zeta=0,\, 0.2,\, {\rm and }\, 0.4$. From left to right: $g_{\uparrow\downarrow}(r)$, $g_{\uparrow\uparrow}(r)$ and $g_{\downarrow\downarrow}(r)$.}
\label{Fig3}
\end{figure}

\begin{figure}
\centerline{\mbox{\psfig{figure=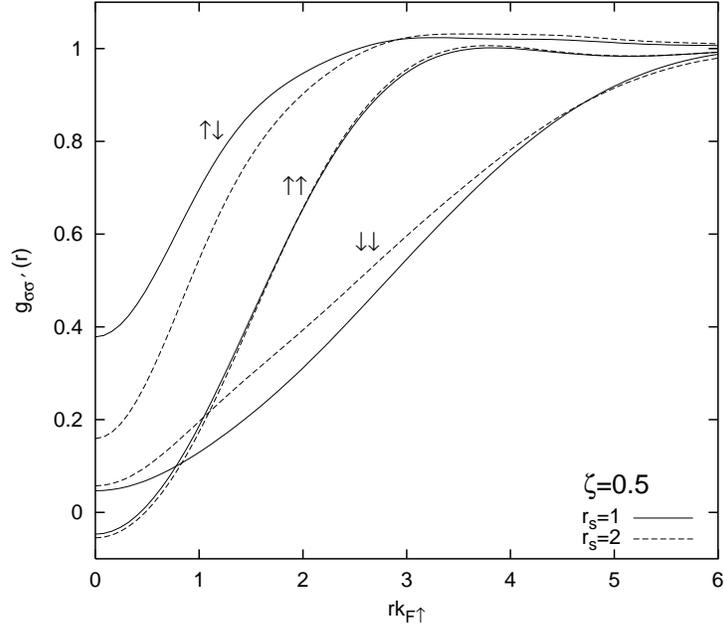, angle =0, width =10 cm}}} 
\caption{
Pair distribution functions {\it versus} $rk_{F\uparrow}$ at $\zeta=0.5$ and $r_s=1\, {\rm and}\, 2$. From left to right: $g_{\uparrow\downarrow}(r)$, $g_{\uparrow\uparrow}(r)$ and $g_{\downarrow\downarrow}(r)$.}
\label{Fig4}
\end{figure}

\begin{figure}
\centerline{\mbox{\psfig{figure=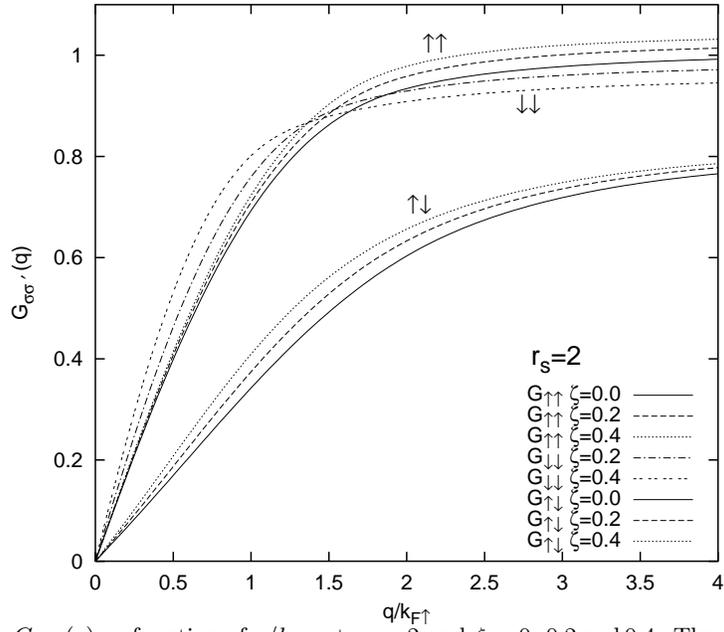, angle =0, width =10 cm}}} 
\caption{Local field factors $G_{\sigma\sigma'}(q)$ as function of $q/k_{F\uparrow}$ at $r_s=2$ and $\zeta=0,\, 0.2\, {\rm and}\, 0.4$. The meaning of the curves is shown in the inset.}
\label{Fig5}
\end{figure}

\begin{figure}
\centerline{\mbox{\psfig{figure=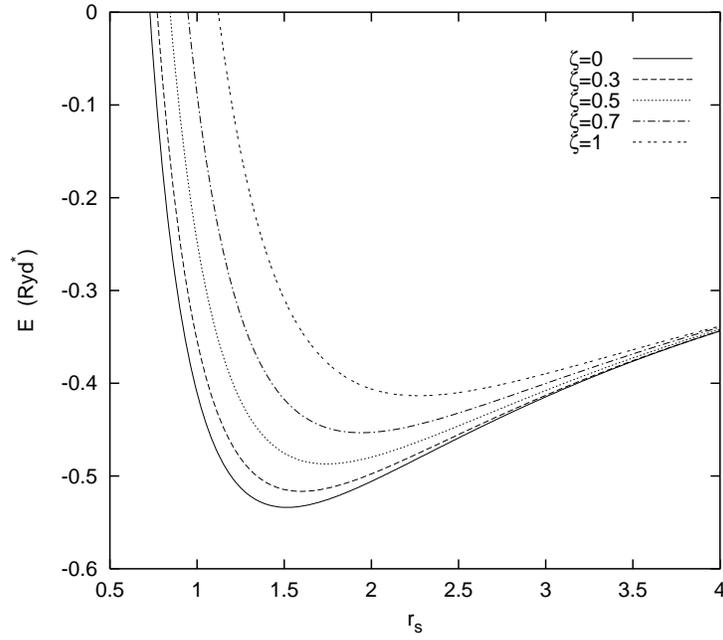, angle =0, width =10 cm}}} 
\caption{Ground state energy as a function of $r_s$ at $\zeta=0,\,0.3,\,0.5,\,0.7\,{\rm and}\,1$.}
\label{Fig6}
\end{figure}

\begin{figure}
\centerline{\mbox{\psfig{figure=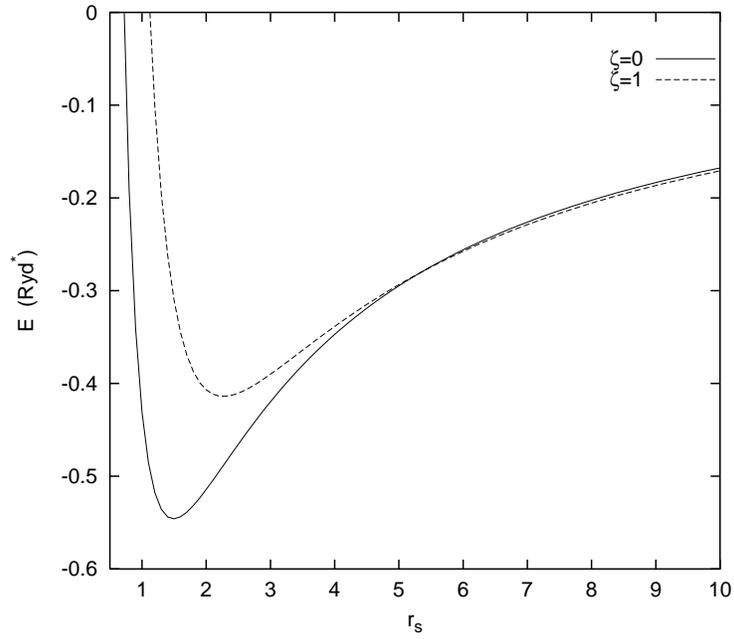, angle =0, width =10 cm}}} 
\caption{Ground state energy as a function of $r_s$ for the paramagnetic and ferromagnetic states, $\zeta=0\,{\rm and}\,1$ respectively.} 
\label{Fig7}
\end{figure}

\begin{figure}
\centerline{\mbox{\psfig{figure=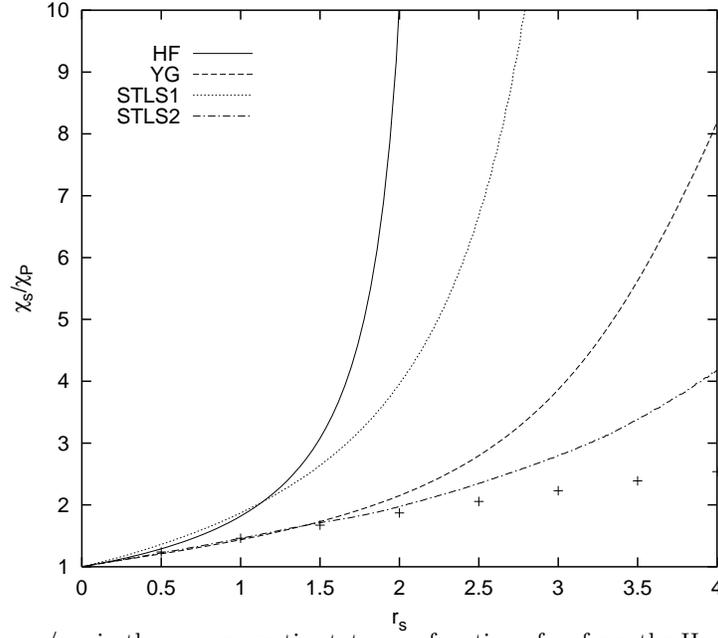, angle =0, width =10 cm}}} 
\caption{Spin susceptibility $\chi_s/\chi_P$ in the paramagnetic state as a function of $r_s$ from the Hartree-Fock approximation (HF), the Yarlagadda-Giuliani calculation (YG), the present calculation according to Eq. (\ref{Xs2}) (STLS1) and to Eq.(\ref{Xs1}) (STLS2), and from Monte Carlo data (crosses)}.
\label{Fig8}
\end{figure}

\begin{figure}
\centerline{\mbox{\psfig{figure=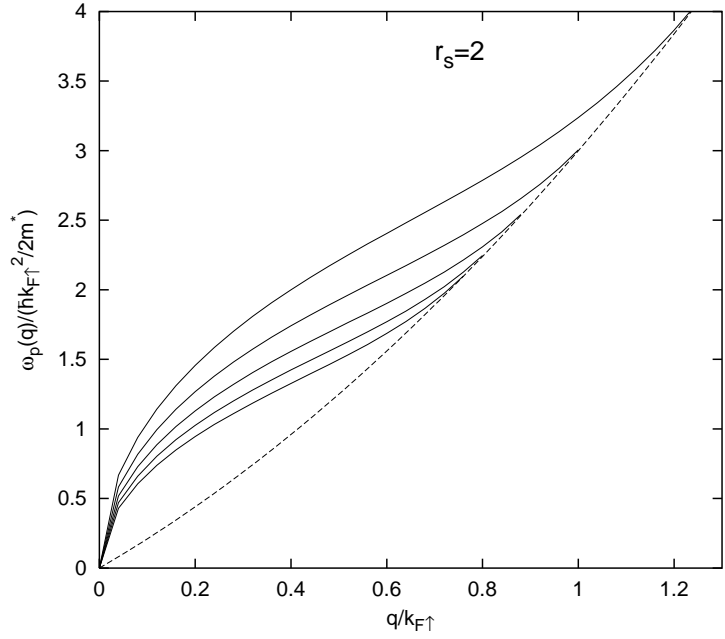, angle =0, width =10 cm}}} 
\caption{Plasmon excitation energy as a function of $q/k_{F\uparrow}$ at $r_s=2$ and  $\zeta=0,\,0.2,\,0.4,\,0.6\, {\rm and}\,0.8$ (from top to bottom).}
\label{Fig9}
\end{figure}

\begin{figure}
\centerline{\mbox{\psfig{figure=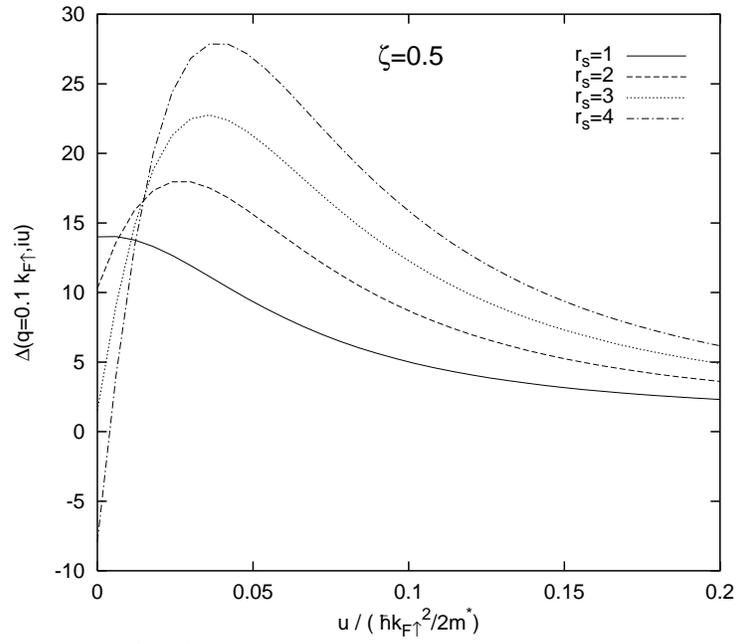, angle =0, width =10 cm}}} 
\caption{$\Delta(q,iu)$ as a function of $u$ at $\zeta=0.5$ and $r_s=1,\,2,\,3\, {\rm and}\, 4$.}
\label{Fig10}
\end{figure}

\end{document}